\newcommand{\nb}{$N$-body } 
\newcommand{\lcdm}{$\Lambda$CDM } 

\newcommand{\lx}{\hbox{$L_\mathrm{X}$}}
\newcommand{\lxtx}{\hbox{$L_\mathrm{X}$-$T$}}
\def\Msun{\hbox{$\rm\, M_{\odot}$}}

\documentclass[final]{aipproc}
\usepackage{amssymb}
\layoutstyle{6x9}

\begin{document}

\title[Galaxy Clusters: the Need for AGN Heating]{Combining Semi-Analytic Models of Galaxy Formation with Simulations of Galaxy Clusters: the Need for AGN Heating}


\author{C.~J.~Short}{address={Astronomy Centre, University of Sussex, Falmer, Brighton, BN1 9QH, United Kingdom}}

\author{P.~A.~Thomas}{address={Astronomy Centre, University of Sussex, Falmer, Brighton, BN1 9QH, United Kingdom}}

\begin{abstract}
We present hydrodynamical \nb simulations of clusters of galaxies with feedback taken from
  semi-analytic models of galaxy formation. The advantage of this technique is that the source of feedback in our simulations is a population of galaxies that closely resembles that found in the real
  universe. We demonstrate that, to achieve the high entropy levels found in clusters,
  active galactic nuclei must inject a large fraction of their energy into the
  intergalactic/intracluster media throughout the growth period of the central black hole.
  These simulations reinforce the argument of \citet{BMB08}, who arrived at the same conclusion
  on the basis of purely semi-analytic reasoning.
\end{abstract}

\maketitle

\section{Introduction}

X-ray observations of groups and clusters of galaxies allow us to probe the physical properties of the hot, diffuse intracluster medium (ICM). In the self-similar collapse scenario, where the ICM is heated solely by gravitational processes, we expect the X-ray luminosity $\lx$ of clusters to scale with gas temperature $T$ as $\lx\propto T^2$. However, the observed \lxtx\ relation is much steeper than predicted, with $\lx\propto T^{2.5-3}$ at $T \gtrsim 2$ keV, becoming steeper still at group scales $T\lesssim 2$ keV. This deficit in X-ray luminosity is due to an excess of entropy \footnote{We define entropy as $S=kT/n_\mathrm{e}^{\gamma-1}$, where $k$ is Boltzmann's constant, $n_\mathrm{e}$ is the electron number density and $\gamma=5/3$ is the ratio of specific heats for a monoatomic ideal gas.} in cluster cores. The source of this excess entropy is likely to be a combination of non-gravitational cooling and heating processes. 

The most obvious sources of non-gravitational heating are Type II supernovae (SNe) and active galactic nuclei (AGN). Cutting-edge hydrodynamical \nb simulations of galaxy groups and clusters attempt to couple cooling, star formation and black hole (BH) growth with associated feedback from SNe and AGN (e.g. \citep{SSD07}). Although in their infancy, such simulations have already yielded encouraging results. For example, \citet{PSS08} resimulated a sample of 21 groups and clusters with the model of \citet{SSD07}, obtaining a mean \lxtx\ relation and halo gas fractions that are in good agreement with observational data over a wide range of mass scales. However, the stellar fraction within the virial radii, $r_{\rm vir}$, of their objects appears larger than observations suggest, implying that too much gas has cooled, even with stellar and AGN feedback.

In this work we pursue a different, but complementary, approach to the theoretical study of galaxy groups and clusters. Instead of undertaking fully self-consistent hydrodynamical simulations, we investigate what current semi-analytic models (SAMs) of galaxy formation predict for the thermodynamical properties of the ICM. These models have already proved capable of explaining many key observational properties of real galaxies. Our goal is to extend the predictive power of SAMs to group and cluster scales by coupling them to cosmological hydrodynamical $N$-body simulations. The main benefit of this hybrid approach is that feedback in our simulations is guaranteed to originate from a realistic galaxy population. This is generally not the case in self-consistent hydrodynamical simulations. 

\section{The numerical model}
\label{sec:method}

Or hybrid technique consists of three components which we now describe in turn. For full details of the modelling process, we refer the reader to \citet{SHT09}.

\paragraph{A dark matter simulation}

We adopt a spatially-flat \lcdm cosmological model with the same parameters as the Millennium simulation \citep{SWJ05}. Initial conditions (ICs) were generated for a cubic volume of side length $L=125h^{-1}$ Mpc containing $N_{\rm DM}=540^3$ dark matter (DM) particles of mass $m_{\rm DM}=8.61\times 10^{8}h^{-1}\Msun$. We evolved these ICs to $z=0$ using the $N$-body/SPH code GADGET-2, storing particle data at the $64$ output redshifts of the Millennium simulation. DM halos are identified on the fly with the friends-of-friends (FOF) algorithm and bound substructures orbiting within FOF halos are found in post-processing with the SUBFIND algorithm. Halo merger trees are then constructed as in \citep{SWJ05}.

\paragraph{Feedback from a semi-analytic model of galaxy formation}

A galaxy catalogue is generated for our DM simulation by applying the L-Galaxies SAM described by \citet{CSW06} and \citet{DLB07} to the merger trees. For each galaxy in the catalogue, we use its merger tree to compute the change in stellar mass, $\Delta M_*$, and mass accreted by the central BH, $\Delta M_{\rm BH}$, between successive model outputs. The energy imparted to the ICM by Type II SNe is then calculated from $\Delta M_*$ via Eq. (20) of \citet{CSW06}. To determine the amount of heat energy input into the ICM by AGN, we do not use the L-Galaxies AGN feedback prescription, but instead use the scheme employed in the version of the SAM GALFORM presented by \citet{BMB08}. The reason for this choice is discussed in \citet{SHT09}. The energy transferred to intracluster gas by AGN then follows from  $\Delta M_{\rm BH}$ by using Eq. (3) of \citet{BMB08}.

\paragraph{A hydrodynamical simulation}

The ICs for our DM simulation are modified by adding gas particles with \emph{zero} gravitational mass. This ensures the DM distribution will be unaffected by the baryons, so the halo merger trees will be the same. A modified version of GADGET-2 is used to evolve these ICs. Whenever an SPH calculation is required, we assign the gas particles their true mass (assuming $\Omega_{\rm b,0}=0.045$), so that gas properties are computed correctly. Once an output redshift is reached, temporary `galaxy' particles are introduced throughout the simulation volume at positions specified by the SAM galaxy catalogue. For each galaxy particle, we know the change in stellar mass and energy released by SNe/AGN since the last output (see above). We use this information to form stars and heat gas in the vicinity of each galaxy as detailed in \citet{SHT09}. The galaxy particles are then removed and the simulation proceeds until the next output time, when the process is repeated. Note that we choose to neglect cooling processes in our simulation (see below). 

\section{Results and discussion}
\label{sec:results}

Two key observables of galaxy groups and clusters are the \lxtx\ relation and the halo gas fraction. We now investigate whether our model can explain the observational data.

\paragraph{The X-ray luminosity-temperature relation}

\begin{figure}
\includegraphics[height=0.35\textheight]{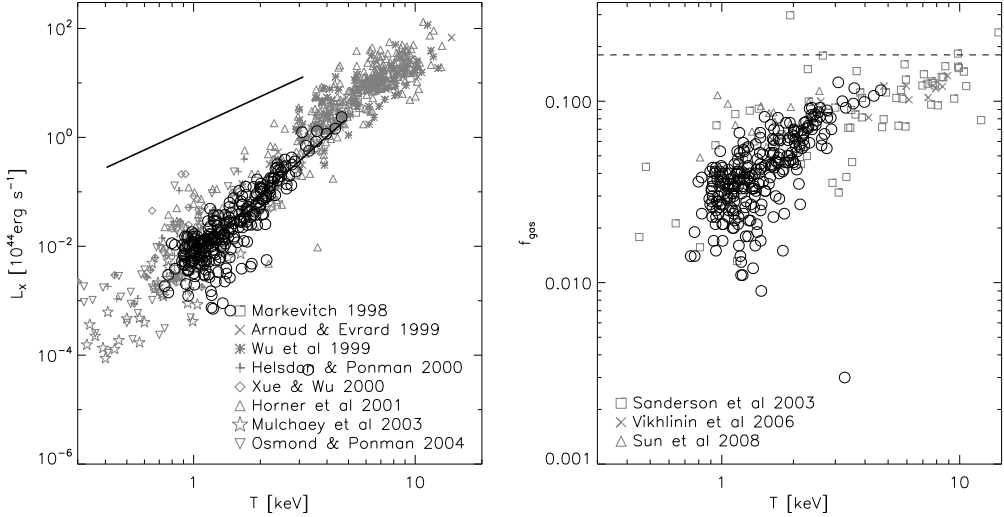}
\caption{Bolometric X-ray luminosity (left panel) and halo gas fraction (right panel) as a function of emission-weighted temperature. All X-ray properties are computed within $r_{500}$. Our simulated groups and clusters are shown by open black circles and several datasets from X-ray observations are also shown for comparison. In the left panel, the upper and lower solid black lines are best fit relations obtained from non-radiative and preheating simulations, respectively. In the right panel, the horizontal dashed line is the mean cosmic baryon fraction for the cosmological model we have adopted.}
\label{fig:plot}
\end{figure}

The \lxtx\ relation obtained from our hybrid simulation at $z=0$ is shown in the left panel of 
Fig. \ref{fig:plot}. It is clear that the slope and normalisation of our relation is consistent with observations at all mass scales. In particular, we see the same steepening on group scales. This is because AGN heating is more efficient at driving X-ray emitting gas from the central regions of low-mass halos, reducing the gas density and thus X-ray luminosity. Our results also exhibit a variation in scatter about the mean relation that is similar to the data for $T\lesssim 3$ keV. Scatter about the
low-temperature end of the \lxtx\ relation is due to the variety of merger
histories of groups. Note that there seems to be more scatter towards the lower-luminosity edge of the observed relation than the upper-luminosity edge. The reason for this is that we cannot produce systems with a highly X-ray luminous cool core since gas cooling is not incorporated in our simulation.

\paragraph{Halo gas fractions}

The halo gas fractions $f_{\rm gas}$ of our groups and clusters are plotted as a function of emission-weighted temperature in the right panel of Fig. \ref{fig:plot}. It is clear that there is a broad agreement with the observational data, with our results exhibiting a comparable amount of scatter. As in the data, we see a rapid decline in gas fraction at lower temperatures. This is because AGN feedback is effective at removing gas from the central
regions of groups since they have a much shallower potential well than massive clusters. In addition, the stellar fraction within $r_{\rm vir}$ is, on average, approximately $9\%$ for our objects, agreeing with observations (e.g. \citep{BPB01,BMBE08}). This is to be expected since star formation in our simulation is driven by a SAM which has been tuned to reproduce the cosmic star formation history. By contrast, self-consistent hydrodynamical simulations typically predict $20-50\%$ of the baryons within $r_{\rm vir}$ are locked-up in stars (e.g. \citep{PSS08,BPB01}).

\section{Summary and conclusion}
\label{sec:conc}

In this work we set out to extend the predictive power of current SAMs of galaxy
formation by investigating the effect of energy feedback from model galaxies on the
properties of intracluster gas. To achieve this objective we have employed a novel hybrid
technique in which a SAM is coupled to a non-radiative hydrodynamical simulation, thus guaranteeing that the source of feedback is a realistic galaxy population. 

Our main conclusion is that a large energy input from AGN (on average, $35\%$ of the available rest mass energy $0.1M_{\rm BH}c^2$) is required over the entire formation history of halos in order to reproduce the observed \lxtx\ relation and halo gas fractions. This supports the conclusion of \citet{BMB08} derived using purely semi-analytic reasoning. Indeed, the relation we have obtained agrees well with observations on all mass scales probed, displaying a similar degree of scatter. This is also true of the halo gas and stellar fractions of our simulated groups and clusters.

In future work we plan to self-consistently incorporate cooling into our hybrid approach. We note that the inclusion of gas cooling could only lead to a decrease in the entropy of intracluster gas, which would reinforce our conclusion with regard to the high degree of AGN feedback that is required to explain the excess entropy found in clusters. 

\bibliographystyle{aipproc} 
\bibliography{bibliography}

\end{document}